\documentclass[twocolumn,aps,prd, preprintnumbers]{revtex4}
\usepackage{latexsym}
\usepackage{tikz}
\usepackage{amsthm}
\usepackage{amsmath}
\usepackage{graphicx}
\usepackage{amssymb}
\usepackage{hyperref}

\usetikzlibrary{decorations.pathmorphing}
\tikzset{snake it/.style={decorate, decoration=snake}}

\newcommand{\gsim}{\lower.7ex\hbox{$\;\stackrel{\textstyle>}{\sim}\;$}}
\newcommand{\lsim}{\lower.7ex\hbox{$\;\stackrel{\textstyle<}{\sim}\;$}}

\newcommand{\be}{\begin{equation}}
\newcommand{\ee}{\end{equation}}
\newcommand{\bea}{\begin{eqnarray}}
\newcommand{\eea}{\end{eqnarray}}

\newcommand{\comment}[1]{}

\newcommand{\bsb}{\boldsymbol}

\def\p{{\bsb p}}
\def\k{{\bsb k}}

\def\x{{\bsb x}}

\def\md{\mathrm{d}}
\def\calR{{\cal R}}

\usepackage{colortbl}
\definecolor{summersky}{cmyk}{0.71,0.33,0,0.14}
\definecolor{flamingo}{cmyk}{0,0.51,0.71,0.14}
\definecolor{rp}{cmyk}{0.2, 1, 0.6, 0}
\definecolor{pacificblue}{cmyk}{0.95,0.3,0, 0.19}
\definecolor{gray60}{cmyk}{0.4,0.4,0,0.8}
\definecolor{green94}{cmyk}{94,0,100,0}
\definecolor{green80}{cmyk}{80,0,90,0}

\begin{document}

\title{Single-Field Consistency relation and $\delta N$-Formalism}
\author{Ali Akbar Abolhasani}
\affiliation{Sharif University of Technology, Tehran, Iran}
\author{Misao Sasaki}
\affiliation{Kavli Institute for the Physics and Mathematics of the Universe (WPI),
Chiba 277-8583, Japan\\
Yukawa Institute for Theoretical Physics, Kyoto University, Kyoto 606-8502, Japan\\
Leung Center for Cosmology and Particle Astrophysics,\\
National Taiwan University, Taipei 10617, Taiwan}

\date{\today}
\preprint{IPMU18-0098, YITP-18-57}

\begin{abstract}
According to the equivalence principal, the long wavelength perturbations must not have any \textit{dynamical} effect on the short 
scale physics up to ${\cal O} (k^2_L/k^2_s)$. Their effect can be always absorbed to a coordinate transformation locally. 
So any physical effect of such a perturbation appears only on scales larger than the scale of the perturbation.
The bispectrum in the squeezed limit of the curvature perturbation in single-field slow-roll inflation is a good example,
where the long wavelength effect is encoded in the spectral index through Maldacena's consistency relation. This implies that
one should be able to derive the bispectrum in the squeezed limit without resorting to the in-in formalism in which
one computes perturbative corrections field-theoretically. 
In this short paper, we show that the $\delta N$ formalism as it is, or more generically the separate universe approach,
when applied carefully can indeed lead to the correct result for the bispectrum in the squeezed limit. Hence 
despite the common belief that the $\delta N$ formalism is incapable of recovering the consistency relation within itself,
it is in fact self-contained and consistent.
\end{abstract}

\maketitle

\section{Introduction}
The $\delta N$ formalism is a powerful tool for calculating the curvature perturbation from inflation without resorting to the standard cosmological 
perturbation theory. The main advantage of this method is that the superhorizon curvature perturbation can be obtained solely by solving 
background homogeneous equations. 
The $\delta N$ formalism in essence is based on the spatial gradient expansion at its leading order, which is called the separate universe approach.
It reveals that any smoothed patch of a perturbed universe on the Hubble horizon scale evolves locally as a homogeneous and isotropic universe.
Thus a long wavelength perturbation on scales much greater than the Hubble horizon size can be treated as a homogeneous perturbation to a fiducial
background universe on each Hubble horizon scale. 
An important consequence of this fact is that superhorizon curvature perturbations are equivalent
to variations of total number of $e$-folds of expansion of background geometry \cite{Starobinsky:1986fxa,Salopek:1988qh,Sasaki:1995aw, Sasaki:1998ug,Lyth:2004gb}. 
There has been an attempt to extend the above idea to anisotropic backgrounds \cite{Abolhasani:2013zya,Talebian-Ashkezari:2016llx}.

Nevertheless, a seemingly drawback of the this formalism shows up when one tries to calculate the bispectrum of the curvature perturbation in 
the simplest inflationary model, i.e., single-field slow-roll inflation. In this case, the non-Gaussianity is local in the sense that the bispectrum is 
dominated by its squeezed limit configuration, and is conveniently characterized by the non-Gaussianity parameter $f_{NL}$, 
which is of the order of the slow-roll parameters and is composed of two terms; one proportional to $\epsilon$ and the other to 
$\eta$ (see below). The former is generally regarded as an intrinsic non-Gaussianity of the inflaton fluctuations due to the
non-linear coupling through gravity. It is then usually delegated to the so-called in-in formalism\cite{Wands:2010af}. 

To clarify our purpose, let us first review the standard implementation of $\delta N$ formalism for calculating the bispectrum of the curvature perturbations. 
The $\delta N$ formula,  ${\cal R}_c=N_{,\phi}\delta\phi_* + N_{,\phi \phi}\delta\phi_* \delta\phi_*+... $ to the second order yields
\begin{align}
&\dfrac{6}{5} f_{ML}(\p)
\nonumber\\
&\quad
\mathop{=}\limits_{p\equiv p_1\approx p_2\gg p_3}
\left[\frac{N_{\phi \phi}}{N_{,\phi}^2}
+\frac{\langle\delta\phi_*(\p)\delta\phi_*(\p-\p_3)\delta\phi_*(\p_3)\rangle}
{2N_{,\phi}P_{\delta\phi_*}(p_3)P_{\delta\phi_*}(p)}
\right]
\nonumber\\
&\qquad\quad=\eta/2+\epsilon\,,
\label{squeezed}
\end{align}
where the first term in the square brackets (or the $\eta/2$ term) is the contribution captured by the conventional $\delta N$ formalism, 
while the second term (or the $\epsilon$ term) is due to the intrinsic non-Gaussianity of the inflaton fluctuations that the $\delta N$ 
formalism \textit{seemingly} fails to address \cite{Wands:2010af}.
However, the equivalence principle tells us that the leading order \textit{physical} effect of a large-scale 
inhomogeneity on the small-scale perturbations is of the order of $(k_l/k_s)^2$, where $k_l$ and $k_s$
are the characteristic wavenumbers of the large-scale and small-scale perturbations, respectively.
This strongly suggests that the above intrinsic non-Gaussianity is not genuinely intrinsic,  but is purely of 
a kinematical origin which can be explained by the $\delta N$ formalism if applied carefully enough.

Our main point is that the delta N formalism, \textit{as it is}, can in fact give the correct result for the non-Gaussianity 
parameter $f_{NL}$, or the bispectrum in the squeezed limit.
Equivalently, one cay say that long wavelength perturbations have no ``dynamical" effect on the short modes, and their only effect is to
modify the homogeneous background from the fiducial background universe. This idea is the essence of the single-field consistency
 relation~\cite{Maldacena:2002vr}.  Hence the contribution of this piece of calculation is to resolve the standard folklore that 
 the $\delta N$ formalism cannot recover the consistency relation because of the missing ``intrinsic" part. 
 In the following, we explicitly show that this contribution can be captured via careful application of the standard $\delta N$ formalism.

\section{\label{Grad-Exp:sec} Gradient Expansion and Separate Universe Approach}
It was shown in \cite{Sasaki:1998ug} that the uniform-$N$ foliation of the spacetime has the unique property 
that the scalar field equations as well as the Einstein equations take the background form for superhorizon perturbations. 
The flat slicing, in which the spatial metric takes the spatially flat form, belongs to this family of foliations on superhorizon scales.
Thus in terms of the local number of $e$-folds $N$ (where \textit{local} means on the scale of the Hubble horizon) as time variable,
\be
N= \int \tilde{H} d\tau
\ee
where $\tilde{H} = \theta (\x,t)/3$ and $d\tau = (1+A) dt$, the Friedmann and the scalar field equations are expressed as \cite{Sasaki:1998ug,Sugiyama:2012tj}
\begin{align}
&\tilde{H}\dfrac{\md}{\md {N}} \left( \tilde{H} \dfrac{\md}{\md {N}} \phi(\x,N) \right)
\cr
&\quad+3 \tilde{H}^2\dfrac{\md}{\md {N}}\phi(\x,N) + V_{,\phi}(\phi(\x,N))=0\,,
\\
&3 M_P^2\tilde{H}^2 = \rho(\x,N)\approx V(\phi(\x,N))\,,
\end{align}
where $M_P$ is the Planck mass.
This shows that a large-scale scalar field perturbation on every smoothed patch of the Hubble horizon scale evolves independently 
as in a homogeneous FLRW universe. This justifies the separate universe approach for this model.

On the other hand, quantum field theory (QFT) tells us the amplitude of quantum vacuum fluctuations. 
The flat slicing is a privileged foliation of spacetime for QFT computations as well because it gives a closed 
form of the action in terms of the scalar field perturbation. In particular, the  amplitude of the $\delta \phi$ 
on a flat slice is easily computed and whose Fourier amplitude is given by $\delta \phi_{\k} = H/\sqrt{2 k^3}$ 
at $k\lesssim Ha$. In real space this can be regarded as a random walk of the inflaton with $\delta\phi=\pm H/(2\pi)$
in one e-folding time on each Hubble horizon patch \cite{Vilenkin:1983xp,Nakao:1988yi}.
Namely,
\be
<\delta \phi_0^2(x,t)>=\left(\frac{H(t)}{2\pi}\right)^2\,,
\ee
where the subscript $0$ on $\delta\phi$ is attached since we have assumed that
the Hubble parameter is space-independent.
However, on scales much larger than the Hubble scale, $H$ may spatially vary $H=H(\x,t)$.
This implies
\be
<\delta \phi^2(x,t)>=\left(\frac{H(\x,t)}{2\pi}\right)^2\,.
\ee

\section{Large scale modulation in $\bsb{\delta\phi}$}

\begin{figure}
\begin{center}
\includegraphics[scale=0.45]{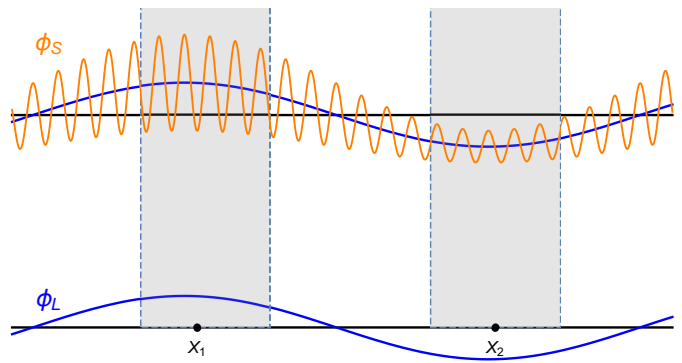}
\end{center}
\caption{This illustration shows the idea that the amplitude of a short wavelength perturbation
	is modulated as a result of an underlying long wavelength perturbation.}
\label{Fig1:SU_idea}
\end{figure}

What we have discussed above has a simple outcome: a short wavelength perturbation
(of $\delta\phi$ on the Hubble horizon scale) is modulated by a long wavelength perturbation
(of the Hubble parameter $H$). 
In the language of the separate universe approach, or the gradient expansion at leading order, 
the amplitude of $\delta\phi$ on the flat slicing at the time when the scale of interest has
just exceeded the horizon size is proportional to the local Hubble parameter $H(\x)$ at that moment.
Thus denoting the fiducial value of $H(\x,t)$ by $\bar{H}(t)$, and the corresponding inflaton perturbation
by $\delta\bar\phi(\x,t)$, the actual inflaton perturbation $\delta\phi(\x,t)$ is modulated as
\be
\label{H-Modulation}
\delta\phi(\x)=\delta\bar\phi(\x)\frac{H(\x)}{\bar{H}},
\ee
where and in what follows we omit the time $t$ from the arguments for notational simplicity.
The idea is shown in Fig. \ref{Fig1:SU_idea}.

Now let us evaluate $H(x)$. At leading order in the slow-roll approximation,
we have $3M_P^2H^2=V(\phi)$. Thus we have
\begin{align}
\frac{H^2(\x)}{\bar{H}^2}=\frac{V(\phi+\delta\phi_L)}{V(\phi)}=1+\frac{V'}{V}\delta\phi_L(\x)+\cdots,
\end{align}
or
\begin{align}
\frac{H(\x)}{\bar{H}}=1+\frac{1}{2}\frac{V'}{V}\delta\phi_L(\x)+\cdots,
\label{deltaH}
\end{align}
where $\delta\phi_L$ indicates that it is the long-wavelength part of $\delta\phi$. 
Inserting the above to Eq. \eqref{H-Modulation} gives
\begin{align}
\delta\phi(\x)=\delta\bar\phi(\x)\left(1+\frac{1}{2}\frac{V'}{V}\delta\phi_L(\x)\right)+\cdots.
\label{deltaphi}
\end{align}

In passing, it may be worth mentioning that the above estimate of $\delta H$ is perfectly consistent
with the result of linear perturbation analysis at long wavelength limit. If we recall the fact that
the curvature perturbation on the flat slicing $\calR_c$ is conserved on superhorizon scales,
and that the gauge transformation from the comoving slicing to the flat slicing on which $\calR=0$ is given by
\begin{align}
0=\calR_c-HT\,,
\end{align}
 hence the lapse function perturbation on the flat slicing $A_f$ on large scales is given by
\begin{align}
A_f=0-\dot T=-\frac{d}{dt}\left(\frac{\calR_c}{H}\right)=-\epsilon\,\calR_c\,,
\end{align}
where $\epsilon\equiv -\dot H/H^2$ and the subscript $f$ is for flat slicing.
Thus one finds
\begin{align}
\frac{\delta H_f}{H}=-A_f=\epsilon\,\calR_c=-\epsilon\,\frac{H}{\dot\phi}\delta\phi_f\,.
\end{align}
We can easily show that this coincides with Eq.~(\ref{deltaH})  in the slow-roll limit.

\section{Revisiting bispectrum computation in the $\bsb{\delta N}$ formalism}

Let us start by a short review of local non-Gaussianity. The so-called local type non-Gaussianity is defined as
\begin{align}
\label{NG-R:def}
{\cal R}_c(\x) = {\cal R}_{cg}(\x) + \dfrac{3}{5} f_{NL} {\cal R}^2_{cg}(\x).
\end{align}
The local type non-Gaussianity predicts a non-vanishing correlation between modes with very different momenta \cite{Babich:2004gb}. 
In particular, the above leads to the bispectrum of the curvature perturbation as
\begin{align}
\langle {\cal R}_c (\k_1) {\cal R}_c (\k_2) {\cal R}_c (\k_3) \rangle 
= +\dfrac{6}{5} f_{NL} {\cal P}_{\cal R} ~ \dfrac{\sum_{i} k_i^3}{\prod_{i} k_i^3}\,,
\end{align}
where ${\cal P}_{\cal R}\propto P_{\cal R}(k)k^3$ is approximately $k$-independent owing to the fact that
 $P_{\cal R} (k)\sim1/k^3$,  and hence the bispectrum is peaked 
at the degenerate triangle of momenta, or the squeezed limit.
 Conversely, in the single-field slow-roll inflation,
 although the bispectrum is non-vanishing for non-degenerate triangles,
 the most important contribution to the bispectrum comes from the squeezed limit which can be well 
 described by the local-type non-Gaussianity, which indicates  that the non-linearities are
 of the superhorizon nature.

According to the above result, the non-linearity of the curvature perturbation in single-field slow-roll inflation 
is significant when there is a correlation between short and long wavelength modes. In real space
this means the most important non-linear term is the product of small and large scale perturbations,  
\be
\label{NG-R:def}
{\cal R}_c(\x) \simeq \bar{{\cal R}}^{s}_{c}(\x) + 2\times \dfrac{3}{5} f_{NL} \bar{{\cal R}}^{s}_{c}(\x) \bar{{\cal R}}^{L}_{c}(\x)+\cdots,
\ee 
when superscripts $\bar{{\cal R}}^{s}$ and $\bar{{\cal R}}^{L}$ refer to  ``Gaussian" perturbations
on small and large length scales, respectively. 
Thus using the $\delta N$ formalism to second order gives 
\begin{align}
\label{deltaN-exp}
{\cal R}_{c} = \delta N
&=\left[\frac{\partial N}{\partial \phi}
+\frac{\partial^2 N}{\partial \phi^2}\delta\phi_L(\x) \right]\delta\phi(\x)
+\cdots.
\end{align}
Rewriting the above in powers of the Gaussian fields by using Eq.~\eqref{deltaphi} yields
\begin{align}
{\cal R}_{c}
&=\frac{\partial N}{\partial \phi}\delta{\bar{\phi}}(\x)
+\left(\frac{\partial N}{\partial \phi}\frac{V'}{2V}
+\frac{\partial^2 N}{\partial \phi^2}\right)
\delta\bar\phi(\x) \delta\phi_L(\x)+\cdots
\cr
&=\bar{{\cal R}}^s(\x)
+\left(\frac{1}{N_{\phi}}\frac{V'}{2V}
+\frac{ N_{\phi \phi}}{N_{\phi}^2}\right)
\bar{\cal R}^s_c(\x) \bar{\cal R}^L_c(\x)+\cdots.
\end{align}
Comparing the above with Eq.~\eqref{NG-R:def}, we find
\be
\dfrac{6}{5} f_{NL} = \frac{1}{N_{\phi}}\frac{V'}{2V}
+\frac{ N_{\phi \phi}}{N_{\phi}^2} = \epsilon + \frac{\eta}{2}\,,
\ee
where in the last step we used $N_{\phi} = V/(M_P^2V') $ and
the definitions of the slow-roll parameters,
\begin{align}
\epsilon&\equiv-\frac{\dot H}{H^2}
=M_P^2\frac{V'{}^2}{2V^2}=\frac{1}{M_P^2N_\phi^2}\,,
\cr
\eta&\equiv\frac{\dot \epsilon}{H\epsilon}=2\frac{N_{\phi\phi}}{N_\phi^2}\,.
\end{align}
Thus we have recovered the total $f_{NL}$ within the $\delta N$ formalism.

Moreover, using the relations,
\begin{align}
\epsilon=\frac{\partial \ln H}{\partial N}\,,
\quad
\eta=2\frac{\partial\phi}{\partial N}\frac{\partial }{\partial\phi}
\ln\frac{\partial N}{\partial\phi}\,,
\end{align}
we obtain
\begin{align}
1-n_s
&=\frac{\partial}{\partial N}\ln
\left[\left(\frac{\partial N}{\partial\phi}\right)^2H^2\right]
\cr
&=\eta + 2\epsilon=\frac{12}{5}f_{NL}\,,
\end{align} 
which is Maldacena's  single-field consistency relation. Thus,  the consistency relation 
may be regarded as another manifestation of the $\delta N$ formula. 
As discussed frequently in the literature, the essence of this relation is that
the three point correlation function of the curvature perturbation in the squeezed limit is 
proportional to the \textit{linear response} of the short wavelength modes to the 
long wavelength mode, while the dynamical effect of the long wavelength fluctuations 
 is suppressed by a factor ${\cal O} (k^2_L/H^2a^2)$ as already mentioned.
 This naturally implies it must be within the scope of
the $\delta N$ formalism, and we have just shown that it is indeed the case.

\section{conclusion}
We showed that careful application of the $\delta N$ formalism recovers
the bispectrum of the curvature perturbation in single-field slow-roll inflation as well
as Maldacena's consistency condition.
 This is in agreement with the fact that  the only effect of large length scale perturbations is kinematical.
  In other words, one can say that Maldacena's consistency relation is a manifestation of
  the $\delta N$ formula.

\begin{acknowledgements}
The authors would like to thanks, Hassan Firouzjahi, Mahdiyar Noorbala and Vincent Vennin for stimulating and fruitful discussions. AAA is very thankful to YITP for the warm hospitality during the long-term workshop
"Gravity and Cosmology 2018" (YITP-T-17-02) during which this work was initiated.   
This work was supported by the JSPS KAKENHI Nos. 15H05888 and 15K21733,
and by World Premier International Research Center Initiative (WPI Initiative), MEXT, Japan.

\end{acknowledgements}

\end{document}